\documentclass[fleqn,twoside,twocolumn,nofootinbib]{revtex4}

\usepackage{graphicx}
\usepackage{dcolumn}
\usepackage{color}
\usepackage{here}
\usepackage{amsthm}
\usepackage{amsmath}
\usepackage{amssymb}
\usepackage{graphicx}
\usepackage{hyperref}
\usepackage{natbib}
\usepackage{titlesec}
\usepackage{array}

\begin{document}
\title{Efficient MPS algorithm for periodic boundary conditions and applications}

\author{Michael Weyrauch}

\affiliation{Physikalisch-Technische Bundesanstalt,\\ Bundesallee
100, D-38116 Braunschweig, Germany}

\author{Mykhailo V. Rakov}
\affiliation{Department of Physics, Taras Shevchenko National
University, \\4 Glushkov av., Kyiv 03127, Ukraine}


\setcounter{page}{1}%

\begin{abstract}
We present an implementation of an efficient algorithm for the
calculation of the spectrum of one-dimensional quantum systems with
periodic boundary conditions. This algorithm is based on a matrix
product representation for quantum states (MPS), and a similar
representation for Hamiltonians and other operators (MPO). It is
significantly more efficient for systems of about 100 sites and more
than for small quantum systems. We apply the formalism to calculate
the ground state and first excited state of a spin-1 Heisenberg ring
and deduce the size of the Haldane gap. The results are compared to
previous high-precision DMRG calculations. Furthermore, we study
spin-1 systems with a biquadratic nearest-neighbor interaction and
show first results of an application to a mesoscopic Hubbard ring of
spinless Fermions which carries a persistent current.
\end{abstract}

\maketitle

\section{Introduction}
It was recognized early on that density matrix renormalization group
(DMRG) simulations of one-dimensional (1D) quantum systems require
significantly more numerical resources for periodic boundary
conditions (PBC) than for open boundary conditions
(OBC)~\cite{WHI93a}. Verstraete, Porras, and Cirac
(VPC)~\cite{VER04} addressed this issue, and they proposed an
algorithm in terms of matrix product states (MPS), which scales
significantly better with the matrix size $m$ of the MPS than
standard DMRG with PBC. However, intermediate steps of this
algorithm require matrices of size $m^2 \times m^2$, and computer
time and memory necessary to determine the improved representation
still scales with $m^5$ as compared to $m^3$ for OBC.

This issue was addressed by Pippan, White, and Evertz
(PWE)~\cite{PIP10}, who recognized that for sufficiently large
systems a much more efficient implementation is possible using a
singular value decomposition (SVD) of products of certain transfer
matrices. In order to calculate such products with sufficient
accuracy only rather few singular values must be kept.

The usefulness of the improved algorithm was demonstrated in
Ref.~\cite{PIP10} by a calculation of the ground state of the spin-1
Heisenberg Hamiltonian. The authors showed that accurate results for
the ground state energy are obtained by a comparison with highly
accurate standard DMRG calculations. As a result, it was concluded
that for large enough systems one obtains an algorithm which scales
similarly with $m$ as calculations for systems with OBC.

In the present paper we extend the PWE algorithm in two respects:
First, we propose an implementation of this algorithm in terms of
MPS and matrix product operators (MPO). To this end we define
generalized transfer matrices, which are subjected to an SVD. This
enables further gains in efficiency in certain situations. Second,
we extend the PWE framework and include the calculation of excited
states of 1D many body Hamiltonians.

We apply this algorithm to a small selection of spin models
(bilinear and biquadratic spin-1), as well as to a spinless Fermion
model. In the course of these applications it was found that in
general the number of singular values one must keep depends on the
matrix size $m$, i.e. the larger $m$ the more singular values must
be kept in order to produce high precision results.

From the MPS representation it is straightforward to calculate
correlation functions and other observables. Results of such
calculations will be presented elsewhere.

\section{MPS-MPO formalism for PBC}

We first rewrite the algorithm proposed in Ref.~\cite{VER04} in
terms of MPS and MPO: The states of a 1D quantum system of size $N$
(e.g. a spin system) are approximated in terms of a matrix product
state (MPS),
\begin{equation}\label{eq:MPS}
|\psi\rangle={\rm Tr}\;B^{[1]}_{\sigma_1}\cdot
\ldots \cdot B^{[N]}_{\sigma_N} |\sigma_1,\ldots,\sigma_N\rangle.
\end{equation}
Here, the $\sigma_j$ represent the local degrees of freedom at site
$j$, and each $B^{[j]}_{\sigma_j}$ represents a matrix of size $m
\times m$, where $m$ is called bond dimension. In the algorithm to
be described the elements of these matrices are variational
parameters to be adjusted using a suitable optimization procedure.
The trace in Eq.~(\ref{eq:MPS}) ensures periodic boundary conditions
and includes a sum over all $\sigma_j$.

Analogously, operators are written as matrix product operators (MPO)
\begin{equation}\label{eq-MPO}
O={\rm Tr}\;W^{[1]}_{\sigma_1,\sigma_1^\prime}  \ldots W^{[N]}_{\sigma_N,\sigma_N^\prime}
|\sigma_1,\ldots,\sigma_N\rangle\langle\sigma_1^\prime,\ldots,\sigma_N^\prime|,
\end{equation}
and the trace includes a sum over all $\sigma_j$ and
$\sigma_j^\prime$. Again, each $W^{[1]}_{\sigma,\sigma^\prime}$
represents a matrix of size $m_W \times m_W$, i.e. each $W$ is a
tensor of order 4. It turns out that all operators of interest with
short range interactions (e.g. the Heisenberg Hamiltonian) can be
written in terms of $W$ tensors with small bond dimensions $m_W$.
The structure of the $W$ tensors is determined by the specific model
under investigation. We will provide the explicit MPO representation
of the various operators later in in this paper.

Matrix elements of MPO in such states,
\begin{equation}\label{eq:matrixe}
\langle\phi|O|\psi\rangle={\rm Tr}\; E_W^{[1]}(A,B)\cdot\ldots \cdot E_W^{[N]}(A,B),
\end{equation}
can be expressed in terms of the (generalized) transfer matrices
\begin{equation}\label{eq-transfer1}
E_W^{[j]}(A,B) = \sum_{\sigma,\sigma^\prime}W^{[j]}_{\sigma\sigma^\prime} \otimes (A^{[j]}_{\sigma})^\star\otimes
 B^{[j]}_{\sigma^\prime}.
\end{equation}
The matrices $A$ and $B$ characterize the states $|\phi\rangle$ and
$|\psi\rangle$, respectively.  The Kronecker product $\otimes$ in
Eq.~\ref{eq-transfer1} obviously produces transfer matrices of size
$ m^2 m_W \times  m^2 m_W$. For later use we also define the special
transfer matrix
\begin{equation}\label{eq-transfer2}
E_1^{[j]}(A,B) = \sum_{\sigma,\sigma^\prime}\delta_{\sigma\sigma^\prime}(A^{[j]}_{\sigma})^\star\otimes
 B^{[j]}_{\sigma^\prime}.
\end{equation}

One advantage of the MPO formalism used here over the formalism
employed by VPC and PWE is the fact that it takes care of the
structure of the effective Hamiltonian to be determined
automatically (as encoded in the MPO), while the effective
Hamiltonian in the VPC formulation depends structurally on the
Hamiltonian of the model under consideration.

In order to find the ground state of the many body system one solves
a standard variational problem using the matrix elements of the MPS
as variational parameters. The optimization of the variational
parameters of the MPS is implemented as a local update step, which
is repeated until convergence is achieved~\cite{VER04}. In the MPO
formalism such a local update step amounts to the solution of a
generalized eigenvalue problem
\begin{equation}\label{eq:eigenproblem-p}
H_{\rm eff}^{[j]}\varphi^{[j]}=\epsilon^{[j]}N_{\rm eff}^{[j]}\varphi^{[j]}
\end{equation}
in terms of the effective Hamiltonian $H_{\rm eff}$ and the
effective normalization matrix $N_{\rm eff}$ given by
\begin{eqnarray}\label{eq:eff-Hamiltonian-p}
H_{\rm eff}^{[j]}&=&\sum_{kl}^{m_W} W^{[j]}_{kl} \otimes \left(\widetilde{H_R^{[j]} \cdot H_L^{[j]}}\right)_{lk},\\
N_{\rm eff}^{[j]}&=&\mathbb{E} \otimes \left(\widetilde{N_R^{[j]}\cdot N_L^{[j]}}\right).
\end{eqnarray}
The energy of the state is obtained from $\epsilon^{[j]}$, and this
value will converge to the ground state energy eventually. In fact,
we stop the iterative update procedure, if this quantity does not
change any more with respect to defined convergence criteria.

The updated MPS is obtained from
$\varphi^{[j]}=M_{(\sigma,l^\prime,l)}$ by a suitable partitioning
of the vector into a tensor. The tilde
in~(\ref{eq:eff-Hamiltonian-p}) indicates the operation
$X_{(ij),(i^\prime j^\prime)}=\tilde{X}_{(ii^\prime),(jj^\prime)}$
for each $m^2 \times m^2$ submatrix of the bracketed quantities. As
a consequence of this transposition the effective Hamiltonian and
the normalization matrix are assured to be Hermitian matrices and
standard methods for the solution of generalized eigenvalue problems
can be applied. (For open boundary conditions the normalization
matrix is unity and only a standard eigenvalue problem needs to be
solved.)

The matrices $H_L^{[j]},~N_L^{[j]}$ and  $H_R^{[j]},~N_R^{[j]} $ are
the products of transfer matrices from all sites to the left and to
the right of the site $j$, where the MPS is updated. The $H$
matrices are obtained from generalized transfer matrices as defined
in Eq.~(\ref{eq-transfer1}), while the $N$ matrices are formed from
the transfer matrices defined in Eq.~(\ref{eq-transfer2}), in both
cases setting $A=B=M$ with  $M$ the MPS to be determined.

In the algorithm proposed by VPC one sweeps back and forth over the
entire lattice several times  updating the MPS at each site until
convergence of the energy $\epsilon^{[j]}$ is achieved. Initially,
one starts from a randomly selected MPS. After each update step the
updated matrix is regauged in order to keep the algorithm stable.
The standard regauging procedure, which assures the relation
\begin{equation}
\sum_{\sigma} B^{[j]}_\sigma  B^{[j]\dagger}_\sigma=1
\end{equation}
after each update step
is described in more detail in Refs.~\cite{PIP10} and~\cite{POR06}.

Similarly, excited states will be constructed iteratively by finding
the lowest state in the space orthogonal to the space spanned by the
states already found. We will denote the matrices of these MPS by
$\Phi_{\sigma,k}^{[j]}$ where $k$ enumerates these states ($k=0$ for
the ground state, $k=1$ for the first excited state, etc.). It was
pointed out in Ref.~\cite{POR06} that this construction can also be
implemented iteratively as an update step by locally projecting to
the orthogonal subspace. Here, we need to determine the local
projection operator $P^{[j]}$ with the property
\begin{equation}\label{eq:proj}
P^{[j]}Y_k^{[j]}=0~~~~~~ \forall~ k
\end{equation}
with
\begin{equation}
Y_k^{[j]}= \widetilde{O_R^{[j]} \cdot O_L^{[j]}} \Phi_k^{[j]}~~~~{\rm and}~~ Y^{[j]\dagger}_k\cdot Y^{[j]}_m=0~~{\rm if}~~k\neq m.
\end{equation}
Here, the spin and $m$ indices of the $\Phi^{[j]}_{\sigma,k}$
matrices are suitably combined to form a vector. For simplicity, we
will use the same symbol $\Phi^{[j]}$ for these vectors (see the
analogous definition of $\phi^{[j]}$ above).

The matrices $O_L^{[j]}$ and $O_R^{[j]}$ are products of transfer
matrices as defined in Eq.~(\ref{eq-transfer2}) from all sites to
the left and to the right of the site $j$, respectively, and setting
$B=M$ and $A=\Phi_k$ with $M$ the (excited) MPS to be determined.
The update procedure for these matrices is implemented as a
generalized eigenvalue problem (see Eq.~(\ref{eq:eigenproblem-p}))
for the projected effective Hamiltonian $P^{[j]}H_{\rm
eff}P^{[j]\dagger}$, and normalization matrices $P^{[j]}N_{\rm
eff}P^{[j]\dagger}$. The (local) projection operator $P^{[j]}$ will
be constructed according to Eq.~(\ref{eq:proj}) by finding a set of
vectors orthogonal to the calculated $Y_k^{[j]}$. A standard
numerical orthogonalization routine is employed for this purpose.

\section{Efficient implementation}

\begin{figure*}[t]
\unitlength1cm
\begin{picture}(4,5)(7.5,0)
 \put(1,0)  {\includegraphics[width=16.5cm]{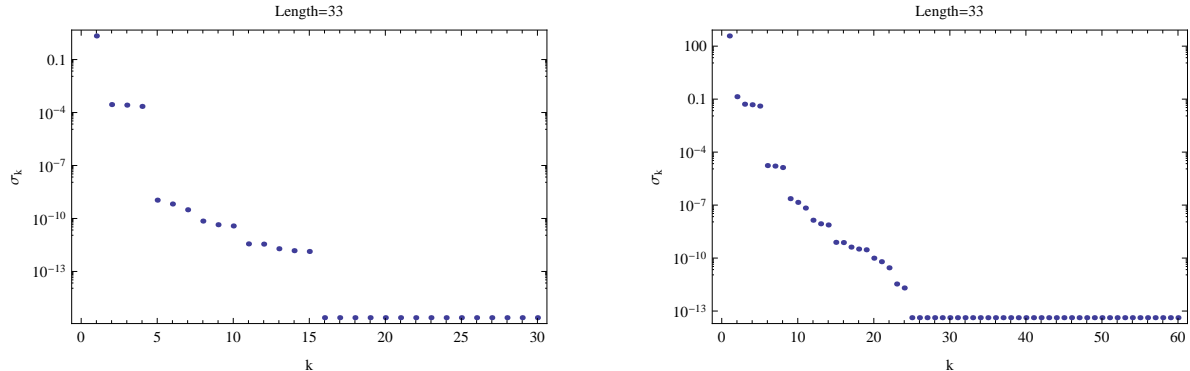}}
\end{picture}
\caption{\footnotesize Distribution of the singular values of
products of transfer matrices for $m$=10, $N_L$ (left) and $H_L$
(right), with 33 terms, which is the minimum number of terms in our
calculation for $N=100$ sites on a ring with homogenous
nearest-neighbor Heisenberg interactions. \label{fig:sing100m10}}
\end{figure*}

In order to implement the local update steps just described one
needs to calculate various products of transfer matrices. These are
standard matrix products, which, however, depending on the bond
dimension of the MPS and MPO, they may be numerically expensive.
Naively, a multiplication of two transfer matrices
(\ref{eq-transfer1}) requires $O(m^6 m_W^3)$ operations, which may
be reduced in view of the structure of the transfer matrices to
$O(m^5 m_W^3)$. In analogy to the proposal by PWE we will now
describe a procedure to reduce this operational count further. This
reduction occurs due to the structure of the $W$ tensors and, in
particular, for products of transfer matrices with many factors,
i.e. long products. Here (unlike Ref.~\cite{PIP10}) we consider
products of transfer matrices in terms of MPSs and MPOs,
\begin{equation}\label{eq:svd}
   E^{[1]}_W(A,B)\cdot\ldots \cdot E_W^{[l]}(A,B)=\sum_{k=1}^{m_W m^2} \sigma_k \;u_k \otimes v_k^\dagger.
\end{equation}
As was pointed out by PWE the sum over $k$ may be cut at rather low
values, which for the generalized transfer matrices has two reasons:
First, the rank $m_S$ of the transfer matrices is in many practical
situations lower than $m_W m^2$. This reduces the upper limit of the
sum to $m_S$. E.g. as is indicated below, the rank of the transfer
matrices for the Ising or Heisenberg models is $2m^2$  and not
$3m^2$ or $5 m^2$, respectively, as expected naively. This reduction
of the summation limit is exact and does not depend on the product
length.

However, for long products, the upper limit may be reduced to very
low values due to the fact that only very few singular values
$\sigma_k$ in the expansion Eq.~(\ref{eq:svd}) are significantly
different from 0. For ground state calculations of chains with about
100 sites and $m=10$ one needs to consider only about 20 singular
values. This is demonstrated for the Heisenberg model in
Fig.~\ref{fig:sing100m10}. This figure corresponds to Fig. 1 of
Ref.~\cite{PIP10} and shows rather similar results for the $N_L$.
Here, we also plot the singular values of $H_L$, and we see that
only a few more singular values than for $N_L$ are needed. (Beyond a
certain limit the singular values are set to an irrelevant small
constant by our computer implementation.)

In order to utilize this feature for the local update algorithm
described in the previous section one needs to implement the
algorithm in such a way, that only sufficiently long products of
transfer matrices occur during the update process. Therefore, one
cannot use the standard sweeping procedure since `short' products of
transfer matrices occur at the turning points of the sweeps.
Following PWE we implement the algorithm as a circular update
procedure. The ring of sites is separated into three sections as
shown in Fig. \ref{fig:circalg}, and the update process occurs
always in the `active' section. The algorithm is then implemented in
3 basic steps:

\begin{enumerate}
\item (Initialization step) Start from some initial randomly created matrix product state
$|\psi\rangle$ as defined in Eq.~\ref{eq:MPS}. The bond
dimension of all matrices ($j=1,\ldots, N$) is $m$. Partition
the set of matrices into three sections as shown in
Fig.~\ref{fig:circalg}.

\begin{figure}
\unitlength1cm
\begin{picture}(6,3)(0,0)
 \put(1,0)  {\includegraphics[width=3.5cm]{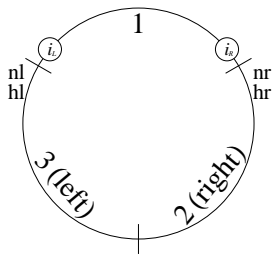}}
\end{picture}
\caption{\footnotesize Circular algorithm for a ring with $N$ sites: the ring is partitioned into three
sections. Updating only happens in one of them, so that we always deal products of transfer matrices with
minimum length $N/3$. For further discussion see the main text.
\label{fig:circalg}}
\end{figure}

Initialize section 3 with a singular value decomposition (SVD)
of the products of generalized transfer matrices defined in
Eqs.~(\ref{eq-transfer1}) and (\ref{eq-transfer2}) and store
this SVD in the tensors \verb"hl" and \verb"nl", respectively.
Initialize section 2 with an SVD of the products of transfer
matrices and store this SVD in the tensors \verb"hr" and
\verb"nr".

\item (Update step) Goto section 1.  Initialize each site of
section 1 with the appropriate product of transfer matrices
moving counter-clockwise starting from the product corresponding
to section 2. Then update and regauge the MPS in section 1
moving clock-wise using the previously calculated products of
transfer matrices. Updating means solving the generalized
eigenvalue problem described above for each site. (One
immediately obtains an SVD of the products of transfer matrices
{\it inside} the updated section by multiplication to the SVD of
the previous site, i.e. one does not need to calculated an SVD
at each update step. This is an important advantage of the
algorithm using MPS and MPO.)

Finally copy the tensors \verb"nl" and \verb"hl" on the tensors
\verb"nr" and \verb"hr" and calculate the SVD of the product of
transfer matrices of section 1 with the just updated MPS
matrices and store this SVD in the tensors \verb"nl" and
\verb"hl". \label{section1}

\item Goto section 2 and do analogous calculations as described
for section 1 above. Continue with further steps moving
clockwise to the neighboring section until convergence is
achieved.
\end{enumerate}


An important prerequisite for the implementation of the algorithm is
an efficient SVD. This has been described in Ref.~\cite{PIP10}, but
we have a few remarks: Let $M$ be a product of transfer matrices.
Then, according to the procedure outlined in Ref.~\cite{PIP10} one
has to form products of these matrices $M$ with some matrices $x$
and $y^\prime$ of size $p\times m^2$, e.g. $y=xM$ and $z=My^{\prime
T}$. In order to do this efficiently one {\it must not} calculate
the matrix $M$ explicitly, but rather multiply each transfer matrix
in $M$ recursively to $x$ or $y^\prime$ starting from one or the
other end of the sequence of factors in $M$. Then the multiplication
of $M$ to the matrices $x$ or $y^\prime$ can be done in $O(Np m^3)$,
where $N$ is the number of transfer matrices in $M$.

Similar steps as outlined above for ground state calculations are
required for the determination of excited states, i.e. for each
excited state we use the same algorithm searching for the optimal
MPS in the space orthogonal to the states already found. We have
implemented the described algorithm within a few pages of {\it
Mathematica} code.

\section{Matrix product operators}

In order to apply the algorithm developed above to specific problems
we must define the relevant degrees of freedom, the size of the
local Hilbert space, and the interaction in terms of a suitable MPO.
Once this MPO is defined, the implementation of the algorithm takes
care of the details of the calculation.

The simplest examples to be considered are spin models, e.g. the
spin-$S$ unisotropic Heisenberg Hamiltonian in an external magnetic
field $B$,
\begin{eqnarray}\label{Heisenberg Hamiltonian}
{\cal H}&=&J\sum_{i=1}^N S_i^x   \otimes S_{i+1}^x+
             S_i^y   \otimes S_{i+1}^y\nonumber\\
   & &  \;         +\Delta S_i^z   \otimes S_{i+1}^z-B\sum_{i=1}^N S_i^z,
\end{eqnarray}
with the exchange interaction $J$, and the unisotropy $\Delta$. In
the following we will set $J=1$. The Hamiltonian is written in terms
of the spin operators $S_i=\frac{1}{2}\sigma_i$, and for
spin-$\frac12$ the $\sigma$ matrices correspond to the standard
Pauli matrices. Periodic boundary conditions correspond to setting
$N+1 \mapsto 1$.

Construction of the MPO for periodic boundary conditions is not
difficult,
\begin{eqnarray}
W^{[1]}&=&\left(
          \begin{array}{ccccc}
            -B S_z & S^x & S^y & S^z & e \\
            0 & 0 & 0 & 0 & S^x \\
            0 & 0 & 0 & 0 & S^y \\
            0 & 0 & 0 & 0 & \Delta S^z \\
            0 & 0 & 0 & 0 & 0 \\
          \end{array}
        \right),\\
W^{[i]}&=&\left(
            \begin{array}{ccccc}
              e & 0 & 0 & 0 & 0 \\
              S^x & 0 & 0 & 0 & 0 \\
              S^y & 0 & 0 & 0 & 0 \\
              S^z & 0 & 0 & 0 & 0 \\
             -B S_z & S^x & S^y & \Delta S^z & e \\
            \end{array}
          \right)\;\;\;\;\mbox{for $i=2,\ldots,N$}\nonumber
\end{eqnarray}
with $e$ a unit matrix. The local single-body Hilbert space has
dimension $2S+1$, and the bond dimension is $d_W=5$. However, the
rank of the transfer matrices which determines the cost of the
calculation is not $5 m^2$ as expected naively but only $2 m^2$. The
first matrix has a different structure as the other matrices in
order to fulfill the required boundary conditions.

For a bilinear-biquadratic spin-$S$ ring with the Hamiltonian
\begin{equation}\label{eq-biqui}
{\cal H}=\sum_{i=1}^N a \vec{S}_i \otimes  \vec{S}_{i+1} +
b (\vec{S}_i \otimes\vec{ S}_{i+1})^2
\end{equation}
one easily finds an explicit MPO representation with a bond
dimension $d_W=14$. Here, again, the rank of the transfer matrices
is not $14 m^2$ but only $2 m^2$, which reduces calculational cost
significantly.

Calculation of matrix elements for observables (e.g. the
magnetization or correlation functions) is straight forward in the
MPS representation either with an MPO representation of the
operators or without. Also for these calculations one may take
advantage of the fact that such calculations are just products of
transfer matrices (see (Eq.~\ref{eq:matrixe})) and use the expansion
(\ref{eq:svd}) for long products. In the present paper we will use
this feature for the calculation of the variance of the Hamiltonian
as is discussed in the next chapter.

\section{Applications}

\begin{table*}[t]
\caption{Ground state energy $E_0$, first excited state energy
$E_1$, and  Haldane gap $E_1-E_0$ for an isotropic  spin-1
Heisenberg ring of $N=100$ sites. $\Delta/E$ is the relative
difference between
our calculated result and the value calculated by DMRG given in Ref.~\cite{PIP10}.\\
\label{tb:res100}} \vskip1mm \tabcolsep10.7pt
\begin{tabular}{|r|l|l|l|l|}
  \hline
  m      &    $E_0/N$     &  $\Delta/E$    &$E_1/N$ (3)      &  $E_1-E_0$   \\
  \hline
  10      & -1.40122726344 & 1.83 $10^{-4}$ & -1.39621210860 & 0.50153      \\
  20      & -1.40145874749 & 1.47 $10^{-5}$ & -1.39730198769 & 0.41566      \\
  30      & -1.40148324293 & 5.83 $10^{-7}$ & -1.39736419879 & 0.41192      \\
  40      & -1.40148390219 & 9.73 $10^{-8}$ & -1.39737237500 & 0.41115      \\
 \hline
  DMRG~\cite{PIP10}
          & -1.4014840386(5) &      -         &   -            &              \\
  DMRG (infinite)~\cite{WHI93a}      & -1.40148403897 &      -         & -1.39737901875   & 0.41050    \\
  \hline
\end{tabular}
\end{table*}

In order to test the implementation of the proposed algorithm we
start out with calculations of the isotropic Heisenberg model also
studied in Ref.~\cite{PIP10}. Of course, it is easily possible to
calculate energy spectra for small systems (10-50 sites) using our
implementation, and we have calculated up to 30 excited states for
such systems. However, then one must take into account most or all
of the singular values in the expansion of the transfer matrices. In
order to take advantage of a significant reduction of the number of
singular values, the system size should be about 100 sites or more,
and we present results for systems with 100 sites in this paper.

In order to run such calculations three important parameters, which
determine the precision of the results must be set: The bond
dimension $m$, the number of singular values to be included in the
expansion of the various transfer matrices $p$ and $p^\prime$, and
the number of update steps $N_u$, where $p$ is the number of
singular values retained in the expansion of the $N_X$ matrices, and
$p^\prime$ those of the $H_X$ matrices.

Of course, a large $m$ is desirable, however, the algorithm scales
at least with $p^\prime m^3 N$, so in practice we are presently
limited to about $m=50$. We shall demonstrate below, that the number
of singular values to be taken into account increases with $m$, and
one must be careful not to take too few terms in the expansion
Eq.~(\ref{eq:svd}). Unfortunately, convergence of the update process
is rather slow close to the minimum of the energy. Therefore, for
high precision results we need more and more update steps, and
usually we choose their number dynamically by observing the change
of the calculated energy within one sector. If this change (averaged
over the whole section) is below a certain limit, we stop the update
process.

One purpose of the present calculations is to gain experience which
parameter setting for $m, p, p^\prime$, and $N_u$ is required  in
order to find e.g. the Haldane gap in a spin-1 ring with sufficient
precision. The gap is obtained from a difference of two large
energies of similar value, so the two energies must be calculated
with rather good precision. (Let us note parenthetically that the
$m$ required in the present algorithm is significantly smaller than
the corresponding quantity in standard DMRG calculations.)

\begin{figure*}[t]
\unitlength1cm
\begin{picture}(4,5)(7.7,0)
 \put(1,0)  {\includegraphics[width=16.5cm]{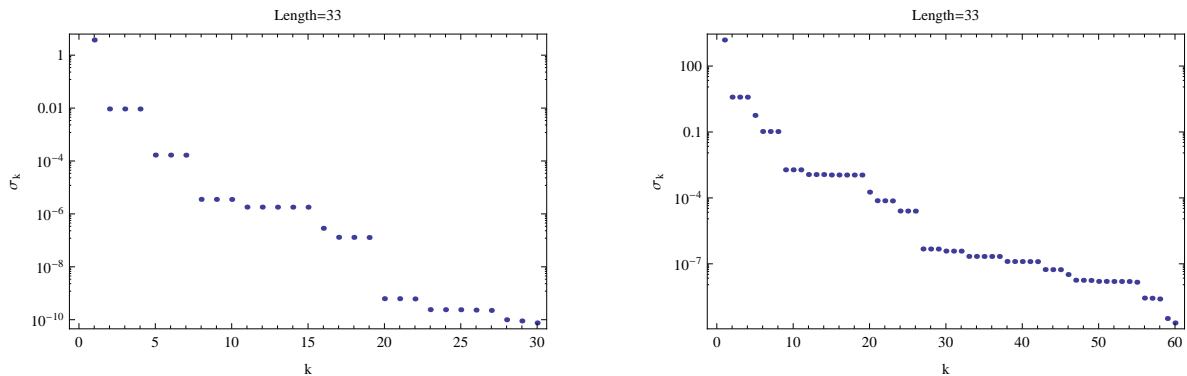}}
\end{picture}
\caption{\footnotesize Distribution of the singular values of
products of transfer matrices for $m$=30, $N_L$ (left) and $H_L$
(right), with 33 terms, which is the minimum number of terms in our
calculation for  $N=100$ sites on a ring with homogenous
nearest-neighbor Heisenberg interactions.\label{fig:sing100m30}}
\vskip .4cm
\end{figure*}

In Fig.~\ref{fig:sing100m30} we show the distribution of singular
values of the transfer matrices obtained at the end of a calculation
with $m=30$ for the isotropic spin-1 Heisenberg model, i.e. the
calculations shown in Fig.~\ref{fig:sing100m10} and
Fig.~\ref{fig:sing100m30} only differ in the choice for $m$. From a
comparison of these results one concludes that if one increases $m$
one also needs to increase the number of singular values to be taken
into account. Our experience shows that the necessary increase is
quite significant depending on the $m$ one wants to use for a
particular calculation. This fact was not mentioned in
Ref.~\cite{PIP10}. However, after this paper was nearly completed,
we became aware that a similar observation was made in
Ref.~\cite{ROS11} for the standard PWE algorithm without MPO.

The MPS-MPO formalism employed here allows to straightforwardly test
how well the calculated MPS approximates an eigenstate of the
Hamiltonian. To this end one calculates the variance
\begin{equation}
\Delta {\cal H} =\langle {\cal H}^2 \rangle - \langle {\cal H} \rangle^2
\end{equation}
which should be zero for an eigenstate. Since from the algorithm we
obtain an explicit representation of the state, we can, at least in
principle, easily evaluate this quantity, if we find a suitable MPO
representation of ${\cal H}^2$. The bond dimension of ${\cal H}^2$
is $m_W^2$, but its rank is often significantly lower, which is used
to significantly reduce the cost for the calculation of ${\cal H}^2$
using the expansion Eq.~(\ref{eq:svd}).

The results obtained so far for the isotropic Heisenberg model are
summarized in Table~\ref{tb:res100}. The ground state energy is in
good agreement with that reported in Ref.~\cite{PIP10}. In addition
we show results for the first excited state from which we determine
the Haldane gap, which agrees with the infinite system DMRG
calculations of Ref.~\cite{WHI93a} to two significant digits.
Haldane~\cite{HAL83} conjectured on the basis of a field theoretical
study that generically integer spin chains are gapped, while
half-integer spin chains are gapless in the thermodynamic limit. For
specific examples (spin-$\frac12$ and spin-$\frac32$) we can confirm
this numerically with our calculations.

For the ground state, we judge the precision of the obtained results
by a comparison to a high precision result obtained by DMRG as
quoted in Ref.~\cite{PIP10}, and assume that this value is
numerically exact for the Heisenberg ring with 100 sites. In fact,
this result is quite close to the infinite system value obtained in
Ref.~\cite{WHI93a}.

A second interesting test of the implementation of the proposed
algorithm is the biquadratic chain Eq.~(\ref{eq-biqui}) (with $a=0$
and $b=-1$) investigated in detail by S\o{}rensen and
Young~\cite{SOR90} using a mapping of the biquadratic spin-1 ring to
the XXZ spin-$\frac{1}{2}$ system, which can be solved using Bethe
Ansatz techniques. In Table~\ref{tb:biqui100} we present some
preliminary results for this system using our technique, which are
compared to the high-precision Bethe Ansatz results of
Ref.~\cite{SOR90}. In the thermodynamic limit one expects a doubly
degenerate ground state and a small gap to the next excited state.
Of course, for finite systems the degeneracy is lifted. This system
is an interesting testing ground for our numerical techniques as
there are extremely precise results available for systems up to 1000
spins. Only for such large systems one expects to be close to the
thermodynamic limit.

The results indicate good agreement with the Bethe Ansatz results,
however, for high precision one needs large $m$ and for $m=30$ one
needs about 30-60 singular values to be taken into account.
Convergence of the energies at a particular $m$, depending on the
precision required, may be slow. Therefore, we recommend to
calculate first with a few different $m$ in order to see the $m$
dependence before one iterates with the chosen $m$ to high
precision. How well the calculated MPS approximates an eigenstate is
measured by a calculation of $\Delta H$.

\begin{table*}[t]
\caption{Ground state energy $E_0$, first excited state energy $E_1$
and gap $E_1-E_0$ for a biquadratic spin-1 Heisenberg ring of
$N=100$ sites ($a=0$, $b=-1$ in Eq.~(\ref{eq-biqui})). $\Delta/E$ is
the relative difference between the Bethe Ansatz results and the
numerical values obtained. ($p=30$, $p^\prime=60$), $\Delta H$ the variance of the Hamiltonian \\
\label{tb:biqui100}} \vskip2mm \tabcolsep10.7pt
\begin{tabular}{|r|l|l|l|l|l|}
  \hline
  $m$                        & $E_0/N$ ( $\Delta/E$ )        & $\Delta H$ & $E_1/N$  ( $\Delta/E$ )  & $\Delta H$   & $E_1-E_0$ \\
  \hline
  10                       & -2.794 020 092 (1.04 $10^{-3}$) & 1.08 & -2.793 830 121 (1.05 $10^{-3}$) & 1.10  &0.018 997   \\
  20                       & -2.795 792 099 (4.07 $10^{-4}$) & 0.44 & -2.795 632 899 (4.12 $10^{-4}$) & 0.44  &0.016 077   \\
  30                       & -2.796 790 186 (5.03 $10^{-5}$) & 0.03 & -2.796 675 842 (3.95 $10^{-5}$) & 0.28  &0.011 452   \\
  \hline
  Bethe Ansatz~\cite{SOR90}& -2.796 930 734 & -                     & -2.796 786 305                 &  & 0.014 442       \\
  \hline
\end{tabular}
\end{table*}

As a last example we apply the proposed algorithm to a Hubbard model
of spinless Fermions, and in particular to a mesoscopic ring filled
with electrons pierced by a magnetic field, such that persistent
currents can be observed. The Hamiltonian of this system is given by
\begin{equation}\label{eq:pc-hamiltonian}
{\cal H}=-t\sum\limits_{\ell=1}^{N}
\left(
c^\dagger_{\ell}c^{}_{\ell+1}e^{-i\phi/N}+{\rm h.c.}
\right)
+U\sum\limits_{\ell=1}^{N}n_{\ell}n_{\ell+1}+V\,n_{1}\, .
\end{equation}
Here $\phi$ is the magnetic flux piercing the ring, $U$ the
nearest-neighbor Coulomb interaction and $V$ the local interaction
of an impurity at site 1. Here, $c^\dagger$ and $c$ are Fermion
creation and destruction operators, and $n$ the density operator.
The hopping energy $t$ will be set to 1, and periodicity requires to
set $N+1  \mapsto 1$.

More details about this Hamiltonian and its physics may be found in
Refs.~\cite{GEN09} and the references therein. The Hamiltonian is
U(1) symmetric, and the particle number is a good quantum number to
label the states. Due to the impurity, the model is not homogeneous:
it is one advantage of our MPS implementation that it can handle
inhomogeneous problems, since it does {\it not} assume translational
invariance of system.

Since we are considering spinless Fermions, the local single-body
Hamiltonian describes a two-level system, which is analogous to a
spin-$\frac12$ system. The matrix representations of the single-body
operators read
\begin{equation}\label{eq:mrep}
c_\ell^\dagger=\left(
            \begin{array}{cc}
              0 & 1 \\
              0 & 0 \\
            \end{array}
          \right),\;\;
c_\ell=\left(
            \begin{array}{cc}
              0 & 0 \\
              1 & 0 \\
            \end{array}
          \right),\;\;
n_\ell=c_\ell^\dagger c_\ell=\left(
            \begin{array}{cc}
              1 & 0 \\
              0 & 0 \\
            \end{array}
          \right).
\end{equation}
Together with the $2 \times 2$ unit matrix these matrices (like the
Pauli matrices) form a complete set.

One then obtains the following MPO representation for this problem,
\begin{eqnarray}
W^{[1]}&=&\left(
          \begin{array}{ccccc}
            (V-\mu) n& -c^\dagger e^{-i\varphi/N} & -c e^{i\varphi/N}& U n & e \\
            0 & 0 & 0 & 0 & c \\
            0 & 0 & 0 & 0 & c^\dagger \\
            0 & 0 & 0 & 0 & n \\
            0 & 0 & 0 & 0 & 0 \\
          \end{array}
        \right),\nonumber\\
W^{[i]}&=&\left(
            \begin{array}{ccccc}
              e & 0 & 0 & 0 & 0 \\
              c e^{i\varphi/N} & 0 & 0 & 0 & 0 \\
              c^\dagger e^{-i\varphi/N} & 0 & 0 & 0 & 0 \\
              n & 0 & 0 & 0 & 0 \\
             -\mu n & c^\dagger & c & Un & e \\
            \end{array}
          \right)\;\;\;\;\mbox{for $i=2,\ldots,N$}\nonumber
\end{eqnarray}
in terms of the matrices defined in Eq.~(\ref{eq:mrep}), the
parameters of the Hamiltonian, and a chemical potential $\mu$  to be
discussed below. The minus signs in the first row of $W^{[1]}$ arise
due to the anti-commutativity of the creation and destruction
operators on different sites.

In order to study persistent currents one needs to calculate the
ground state energy as a function of the magnetic flux and then
calculate the persistent current $j$ using the Hellmann-Feynman
theorem, $j=-{\partial E_0(\phi)}/{\partial \phi}$.

Since experiments are usually made for systems with fixed particle
number, it would be necessary to develop the algorithm in such way
that it respects the U(1) symmetry of the Hamiltonian. At this stage
our implementation does not respect this symmetry. Of course, it is
always possible to shift to the ground state of the sector with the
desired particle number using an appropriate chemical potential
$\mu$. However, this chemical potential is usually not known, and
one would need to use an iteration process to find that chemical
potential such that the resulting state contains the desired number
of particles. Only for half-filled systems, it is known that the
required chemical potential to find the ground state equals the
interaction $U$. We therefore concentrate here on half-filled
systems, and shift the spectrum accordingly.

First results are shown in Table~\ref{pcurr} for a ring with $N=128$
sites. In order to be able to calculate persistent currents using
Hellmann-Feynman theorem one must be able to precisely distinguish
the ground state energies for different $\phi$, which requires
rather high-precision calculations. The energy determined for the
ground state agrees with the result given in Ref.~\cite{GEN09}. We
also calculate the energy of the next higher/lower state and the
number of particles $n$ it contains. Clearly, the ground state is
half-filled, while the next higher/lower state contains one particle
less. At $\phi=0$ one finds a degenerate ground state in the
half-filled sector. (Here, our procedure to calculate `excited'
states, may yield even a lower lying state, since within the
spectrum there exist states below the ground state of the half
filled sector.) For future calculations an implementation respecting
the U(1) symmetry is desirable. 

\begin{table}[h]
\caption{Energy of the half-filled ground state $E_0$ and energy of
the next higher/lower  state $E_1$ of a spinless Fermion ring
described by the Hamiltonian Eq.~(\ref{eq:pc-hamiltonian}) for
$N=128$, $m=30$, $U=1$, and $V=0$. \label{pcurr}}\vskip3mm
\tabcolsep8.7pt
\begin{tabular}{|l|l|l|l|l|}
  \hline
    $\phi$ &   $E_0$       & $n$  & $E_1$         & $n$    \\
  \hline
  $0$      & -63.98647233  & 64 & -63.98581164  & 64       \\
  $\pi/2$  & -64.00411240  & 64 & -64.94361781  & 63       \\
  $\pi$    & -64.01004832  & 64 & -64.94770847  & 63       \\
  \hline
\end{tabular}
\end{table}

\vskip .2cm

\section{Summary}

In this paper we suggest a new version of an efficient MPS algorithm
for one dimensional systems with periodic boundary conditions. The
present version unlike the original proposal~\cite{PIP10} uses an
MPO representation. We also extend the algorithm for the calculation
of excited states. We report about first results obtained with this
algorithm, and investigate the necessary parameter settings in order
to obtain high precision results for systems with 100 sites. The
advantage of the algorithm is that one obtains an explicit
representation of the many-body quantum state, which can then be
used to calculate observables such as correlation functions. We will
report about such calculations in a forthcoming publication.

\vskip3mm

We thank H. G. Evertz for a helpful correspondence. M. V. Rakov
thanks Physikalisch-Technische Bundesanstalt for financial support
during three short visits to Braunschweig.


\begin{thebibliography}{8}
\expandafter\ifx\csname
natexlab\endcsname\relax\def\natexlab#1{#1}\fi
\expandafter\ifx\csname bibnamefont\endcsname\relax
  \def\bibnamefont#1{#1}\fi
\expandafter\ifx\csname bibfnamefont\endcsname\relax
  \def\bibfnamefont#1{#1}\fi
\expandafter\ifx\csname citenamefont\endcsname\relax
  \def\citenamefont#1{#1}\fi
\expandafter\ifx\csname url\endcsname\relax
  \def\url#1{\texttt{#1}}\fi
\expandafter\ifx\csname urlprefix\endcsname\relax\def\urlprefix{URL
}\fi \providecommand{\bibinfo}[2]{#2}
\providecommand{\eprint}[2][]{\url{#2}}

\bibitem[{\citenamefont{White and Huse}(1993)}]{WHI93a}
\bibinfo{author}{\bibfnamefont{S.~R.} \bibnamefont{White}} \bibnamefont{and}
  \bibinfo{author}{\bibfnamefont{D.~A.} \bibnamefont{Huse}},
  \bibinfo{journal}{Phys. Rev. B} \textbf{\bibinfo{volume}{48}},
  \bibinfo{pages}{3844} (\bibinfo{year}{1993}).

\bibitem[{\citenamefont{Verstraete et~al.}(2004)\citenamefont{Verstraete,
  Porras, and Cirac}}]{VER04}
\bibinfo{author}{\bibfnamefont{F.}~\bibnamefont{Verstraete}},
  \bibinfo{author}{\bibfnamefont{D.}~\bibnamefont{Porras}}, \bibnamefont{and}
  \bibinfo{author}{\bibfnamefont{J.~I.} \bibnamefont{Cirac}},
  \bibinfo{journal}{Phys. Rev. Lett.} \textbf{\bibinfo{volume}{93}},
  \bibinfo{pages}{227205} (\bibinfo{year}{2004}).

\bibitem[{\citenamefont{Pippan et~al.}(2010)\citenamefont{Pippan, White, and
  Evertz}}]{PIP10}
\bibinfo{author}{\bibfnamefont{P.}~\bibnamefont{Pippan}},
  \bibinfo{author}{\bibfnamefont{S.~R.} \bibnamefont{White}}, \bibnamefont{and}
  \bibinfo{author}{\bibfnamefont{H.~G.} \bibnamefont{Evertz}},
  \bibinfo{journal}{Phys. Rev. B} \textbf{\bibinfo{volume}{81}},
  \bibinfo{pages}{081103R} (\bibinfo{year}{2010}).

\bibitem[{\citenamefont{Porras et~al.}(2006)\citenamefont{Porras, Verstraete,
  and Cirac}}]{POR06}
\bibinfo{author}{\bibfnamefont{D.}~\bibnamefont{Porras}},
  \bibinfo{author}{\bibfnamefont{F.}~\bibnamefont{Verstraete}},
  \bibnamefont{and} \bibinfo{author}{\bibfnamefont{J.~I.} \bibnamefont{Cirac}},
  \bibinfo{journal}{Phys. Rev. B} \textbf{\bibinfo{volume}{73}},
  \bibinfo{pages}{014410} (\bibinfo{year}{2006}).

\bibitem[{\citenamefont{Rossini et~al.}(2011)\citenamefont{Rossini, Giovanetti,
  and Fazio}}]{ROS11}
\bibinfo{author}{\bibfnamefont{D.}~\bibnamefont{Rossini}},
  \bibinfo{author}{\bibfnamefont{V.}~\bibnamefont{Giovanetti}},
  \bibnamefont{and} \bibinfo{author}{\bibfnamefont{R.}~\bibnamefont{Fazio}},
  \bibinfo{journal}{J. Stat. Mech.}  \bibinfo{pages}{P05021}
  (\bibinfo{year}{2011}).

\bibitem[{\citenamefont{Haldane}(1983)}]{HAL83}
\bibinfo{author}{\bibfnamefont{F.~D.~M.} \bibnamefont{Haldane}},
  \bibinfo{journal}{Phys. Lett.} \textbf{\bibinfo{volume}{93A}},
  \bibinfo{pages}{464} (\bibinfo{year}{1983}).

\bibitem[{\citenamefont{S\o{}rensen and Young}(1990)}]{SOR90}
\bibinfo{author}{\bibfnamefont{E.~S.} \bibnamefont{S\o{}rensen}}
  \bibnamefont{and} \bibinfo{author}{\bibfnamefont{A.~P.} \bibnamefont{Young}},
  \bibinfo{journal}{Phys. Rev. B} \textbf{\bibinfo{volume}{42}},
  \bibinfo{pages}{754} (\bibinfo{year}{1990}).

\bibitem[{\citenamefont{Gendiar et~al.}(2009)\citenamefont{Gendiar, Krcmar, and
  Weyrauch}}]{GEN09}
\bibinfo{author}{\bibfnamefont{A.}~\bibnamefont{Gendiar}},
  \bibinfo{author}{\bibfnamefont{R.}~\bibnamefont{Krcmar}}, \bibnamefont{and}
  \bibinfo{author}{\bibfnamefont{M.}~\bibnamefont{Weyrauch}},
  \bibinfo{journal}{Phys. Rev. B} \textbf{\bibinfo{volume}{79}},
  \bibinfo{pages}{205118} (\bibinfo{year}{2009}).

\end{thebibliography}

%
%
%

\end{document}